%% file: scit.tex
\documentclass[
aps,
prl,
superscriptaddress,
12pt,
reprint,
nofootinbib,
floatfix
]{revtex4-1}

\usepackage[utf8]{inputenc}
\usepackage[T1]{fontenc}
\usepackage[english]{babel}

\usepackage{graphicx}

\usepackage{amsmath}
\usepackage{amssymb}
\usepackage{mathtools}

\usepackage[colorlinks=true, allcolors=blue]{hyperref}
\usepackage[all]{hypcap}  % for images: anchor the image, not the caption

\usepackage{braket}
\usepackage{xcolor}
\usepackage[normalem]{ulem}

\renewcommand{\vec}[1]{\boldsymbol{#1}}

\begin{document}
\title{Synthetic dimension-induced conical intersections in Rydberg molecules}
\date{\today}
\author{Frederic Hummel}
\email{frederic.hummel@physnet.uni-hamburg.de}
\affiliation{Zentrum für Optische Quantentechnologien, Fachbereich Physik, Universität Hamburg, Luruper Chaussee 149, 22761 Hamburg, Germany}
\author{Matthew T. Eiles}
\affiliation{Max-Planck-Institut für Physik komplexer System, Nöthnitzer Str. 38, 01187 Dresden, Germany}
\author{Peter Schmelcher}
\affiliation{Zentrum für Optische Quantentechnologien, Fachbereich Physik, Universität Hamburg, Luruper Chaussee 149, 22761 Hamburg, Germany}
\affiliation{The Hamburg Centre for Ultrafast Imaging, Universität Hamburg, Luruper Chaussee 149, 22761 Hamburg, Germany}

\begin{abstract}
We observe a series of conical intersections in the potential energy curves governing both the collision between a Rydberg atom and a ground-state atom and the structure of Rydberg molecules.
By employing the electronic energy of the Rydberg atom as a synthetic dimension we circumvent the von Neumann-Wigner theorem.
These conical intersections can occur when the Rydberg atom’s quantum defect is similar in size to the electron-–ground-state atom scattering phase shift divided by $\pi$, a condition satisfied in several commonly studied atomic species.
The conical intersections have an observable consequence in the rate of ultracold $l$-changing collisions of the type Rb$(nf)$+Rb$(5s)\to$ Rb$(nl>3)$+Rb$(5s)$. In the vicinity of a conical intersection, this rate is strongly suppressed, and the Rydberg atom becomes nearly transparent to the ground-state atom.
\end{abstract}

\maketitle

The Born-Oppenheimer approximation is a cornerstone of chemical and molecular physics. 
It provides us with the adiabatic separation of the fast electronic from the slow vibrational motion, resulting in adiabatic potential energy surfaces (PES) determined by the electronic structure for a given nuclear geometry \cite{Born1927}. 
When PES become degenerate at a conical intersection (CI), non-adiabatic interaction effects are important and the Born-Oppenheimer-based intuition developed in molecular physics breaks down \cite{Koppel2007,Worth2004,Worth2008}. CI occur frequently in larger molecules with many vibrational degrees of freedom, for example in the nucleobases \cite{Barbatti2010}, and play a key role in photosynthesis \cite{Hammarstrom2008}. 
CI are responsible for ultrafast radiationless decay mechanisms on the femtosecond time scale \cite{Koppel2007,Arnold2018,Mabrouk2020}.
A controlled environment to study CI can be provided by external optical fields \cite{Moiseyev2008,Sindelka2011} or by ultracold interacting Rydberg systems \cite{Wuster2011,Gambetta2020}.
Diatomic molecules provide here an exception.
Generally, the von Neumann-Wigner non-crossing theorem forbids the crossing of potential energy curves (PEC) in diatomic systems, which are determined by a single vibrational parameter, the internuclear coordinate $R$ \cite{vonNeumann1993}. 

Excited electronic states are key players in the appearance of CI and a particular diatomic system, the collisional complex consisting of a Rydberg atom and a ground-state atom, has attracted significant interest since nearly the advent of quantum physics \cite{Fermi1934,Omont1977,Hickman1979,Gallagher1988,Lebedev1996}.
The interaction between these two atoms, as described by the Fermi pseudopotential, is primarily determined by the $s$-wave  electron-atom scattering phase shift $\delta_s$ \cite{Fermi1934,Omont1977}.
The resulting Born-Oppenheimer PEC can be labeled by the principal quantum number $n$ and angular momentum quantum number $l$ of the Rydberg atom. 
These PEC can be sufficiently attractive to support bound states, known as ultra-long-range Rydberg molecules \cite{Greene2000,Bendkowsky2009,Fey2019rev,Eiles2019}.
At positive energies, they are responsible for the collisional dynamics between the two highly asymmetric atomic partners.
Just as the characteristics of a Rydberg atom are smoothly varying polynomial functions of $n$, so too are the typical molecular properties, for example the potential depths, dipole moments, and bond lengths. 

For this reason, it is illustrative to imagine that this diatomic system evolves along a two-dimensional potential energy \emph{surface}, where the principal quantum number $n$ plays the role of an additional \emph{synthetic} dimension.
In many other contexts, the introduction of synthetic dimensions provides a means to control the dimensionality of a system by mimicking additional degrees of freedom. 
Synthetic dimensions have been realized in optical lattices \cite{Levi2011,Kolkowitz2017}, and have applications in the study of gauge fields \cite{Celi2014}, quantum simulation \cite{Boada2012}, and photonics \cite{Ozawa2019}.

In the present work, we utilize the synthetic dimension $n$ to circumvent the non-crossing theorem, and in doing so, we show that CI can appear in the Rydberg--ground-state atom PES.
These CI occur when the Rydberg atom is initially in a state $\ket{nl_0}$ whose fractional quantum defect $\mu_{l_0}$ is similar in size to $\delta_s/\pi$, which is typically satisfied in states with more than $d$-wave angular momentum $l_0$. 
We highlight results for Rb, where $l_0 = 3$, since interest in such a state has recently grown due to its importance for the preparation of circular Rydberg states in the scope of quantum simulation and quantum computing based on neutral-atom platforms \cite{Anderson2013,Nguyen2018,Cortinas2020,Cantat2020}.

Similar to how CI in more traditional molecular systems provide fast radiationless decays between electronic states, the synthetic CI can facilitate fast dynamics in Rydberg--ground-state atom collisions. 
In particular, we demonstrate that they dramatically suppress the $l$-changing collision rate, which is otherwise a dominant process in this collision. 
This provides a clear experimental observable, heralding the presence of synthetic CI in several different atomic species.

\begin{figure}
    \centering
	\includegraphics[width=0.48\textwidth]{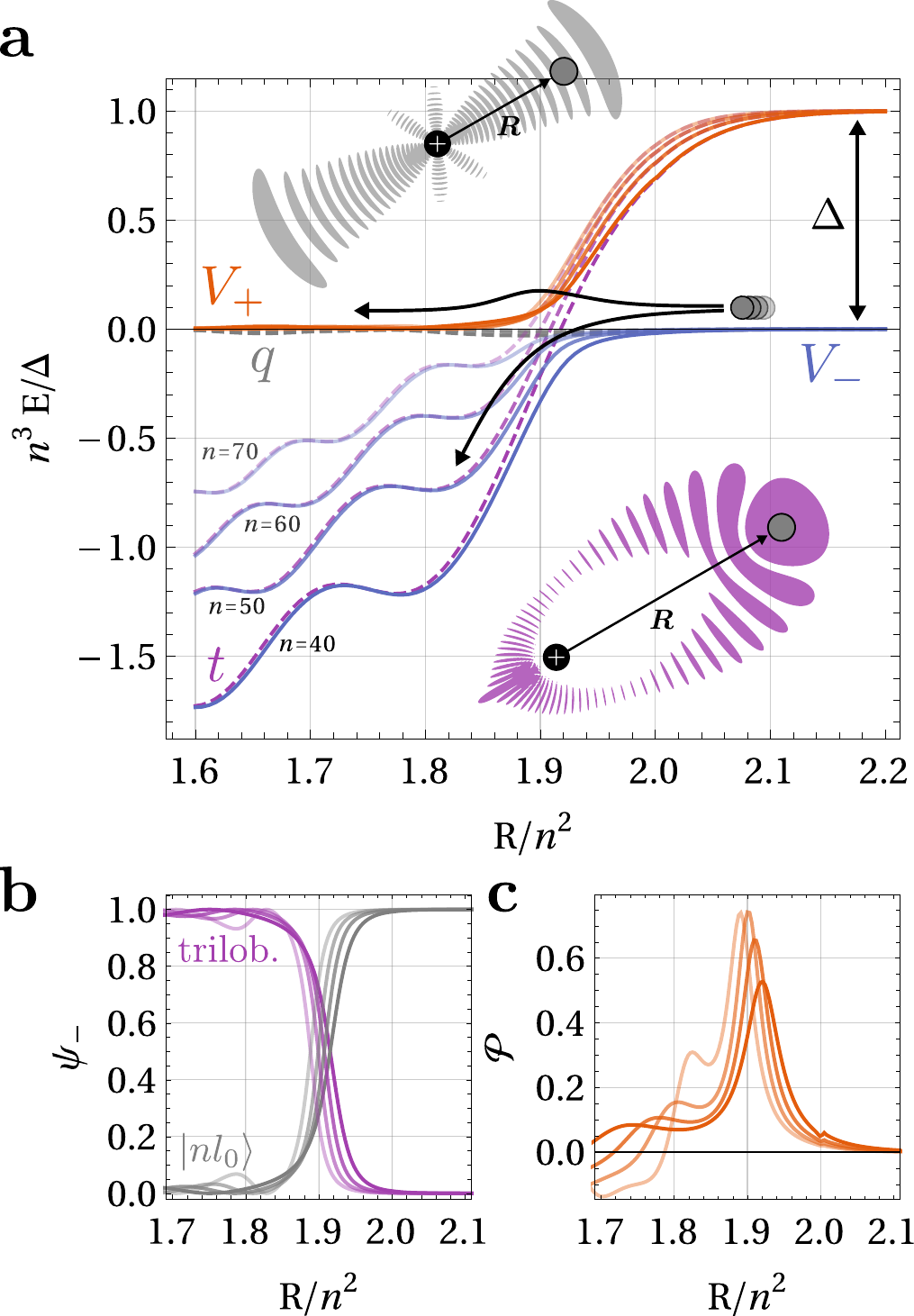}
	\caption{(a) Illustration of the atomic collision and the corresponding adiabatic (solid, red $V_+$ and blue $V_-$) and diabatic (dashed, purple $t$ and gray $q$) PEC for different principle quantum numbers $n\in\{40,50,60,70\}$ (increasing $n$ is plotted with increasing opacity). 
	At large internuclear distances, the difference of the potentials $\Delta$ is determined by the quantum defect of the state with angular momentum $l_0$. 
	The axes are rescaled in order to shift the curves onto the same energy scale. A semi-classical ground-state atom propagating on $V_-$ can either adiabatically follow the potential and pick up orbital angular momentum to form a trilobite state, the electronic density of which is shown at the bottom, or hop to the potential $V_+$ and continue with an $f$-state electronic orbital depicted at the top.
	(b) The eigenvector components $\psi_-$ corresponding to $V_-$ and (c) the non-adiabatic coupling matrix element $\mathcal{P}$ with varying (scaled) $R$ for different $n$.
	\label{fig:scheme}}
\end{figure}

The interaction between the two atoms is given by Fermi's pseudopotential,
\begin{equation}
    V(\vec r, \vec R) = 2\pi a(k)\delta^3(\vec r-\vec R),
\end{equation}
where $a(k) = \frac{\tan\delta_s(k)}{k} =a(0)+\frac{\pi\alpha}{3}k$ is the low-energy $s$-wave scattering length, $\alpha$ is the ground-state atoms's polarizability, and the wave number $k$ is determined semiclassically by $\frac{k^2}{2}-\frac{1}{R}=-\frac{1}{2n^2}$. 
Higher partial waves can be neglected because the electron energy is close to zero at the large internuclear distances considered here.
Since $V(\vec r, \vec R)$ is a weak perturbation to the Coulomb potential, we calculate the PEC by diagonalizing $V(\vec r,\vec R)$ within a restricted basis. This includes only the initial state $\ket{nl_0}$, whose energy we take as the reference energy, and the degenerate manifold of hydrogenic $l>l_0$ states with vanishingly small quantum defects. These are blue-detuned by a shift $\Delta$ from the initial state.  The $l<l_0$ states can be ignored since they are far detuned, $\Delta_{l<l_0}\gg \Delta$; those states with $l>l_0$ have very small detunings relative to $\Delta$, and can be subsumed into the degenerate manifold. 

Two adiabatic PEC result from this diagonalization,
\begin{equation}
    V_\pm(n,R) = \frac{1}{2}\left(q+t+\Delta \pm \sqrt{(q+t+\Delta)^2-4q\Delta}\right), \label{eq:V}
\end{equation}
where 
\begin{subequations}
\begin{align}
    q(n,R) &= a(k) \frac{2l_0+1}{2} \frac{u_{nl_0}^2(R)}{R^2},\\
    t(n,R) &= a(k) \sum_{l=l_0+1}^{n-1} \frac{2l+1}{2} \frac{u_{nl}^2(R)}{R^2}, 
\end{align} 
\end{subequations}
and the hydrogen atom's radial wave functions are $u_{nl}(r)$.
The accuracy of the PEC are confirmed quantitatively by large-scale diagonalizations involving many more Rydberg states \cite{Greene2000,Hamilton2002}, by Green's function methods \cite{Chibisov2002,Khuskivadze2002,Greene2006}, or via local frame transformation techniques \cite{Giannakeas2020a}.

The PEC are shown in Fig.~\ref{fig:scheme}a using scaled distances and energies for several principle quantum numbers $n\in\{40,50,60,70\}$. 
They represent the key ingredient for the investigation of the collisional dynamics between the Rydberg atom $\ket{nl_0}$ and a ground-state atom, which is initialized at large $R$ with the ground-state atom propagating freely with velocity $v$ along the asymptotically flat potential $V_-$. 
The degenerate manifold is strongly perturbed by the ground-state atom, causing an attractive potential $V_+$ to descend. 
At large $R$, its adiabatic electronic state $\psi_+$ is a superposition of high-$l$ states known colloquially as the \emph{trilobite} state (see inset wave function). 
Level repulsion induces an avoided crossing at $R = R_0 \approx 2n^2$, and as a result $V_-$ sharply bends away.

From this picture, we see that, if the collision is purely adiabatic, the Rydberg atom's electronic state will change from its original electronic state (see upper inset) into the trilobite state (see lower inset) as the ground-state atom drops into the attractive potential of $V_-$. 
This is illustrated in figure \ref{fig:scheme}b, which shows how the components $t$ and $q$ of the corresponding adiabatic state $\psi_-$ reversed after the crossing is traversed. 
This has observable consequences, as it induces an $l$-changing collision at a rate $\gamma = \sigma\rho v$, where  $\sigma$ is the collisional cross section and $\rho$ the density of ground-state atoms \cite{Niederprum2015,Gallagher1994}.
Since the Rydberg state changes character nearly at the outer turning point of the Rydberg orbit due to the steep potential curve arising from the strong perturbation of the degenerate manifold, we take $\sigma = \pi R_0^2$. 
This purely adiabatic rate $\gamma$ holds unless the velocity or the non-adiabatic coupling are large enough for a non-adiabatic transition to occur, in which case the ground-state atom hops across onto $V_+$, preserving the electronic character of the Rydberg atom which continues through diabatically. 
In particular, at a CI, where $V_{\pm}$ become degenerate, the non-adiabatic couplings diverge.

\begin{figure}
    \centering
	\includegraphics[width=0.47\textwidth]{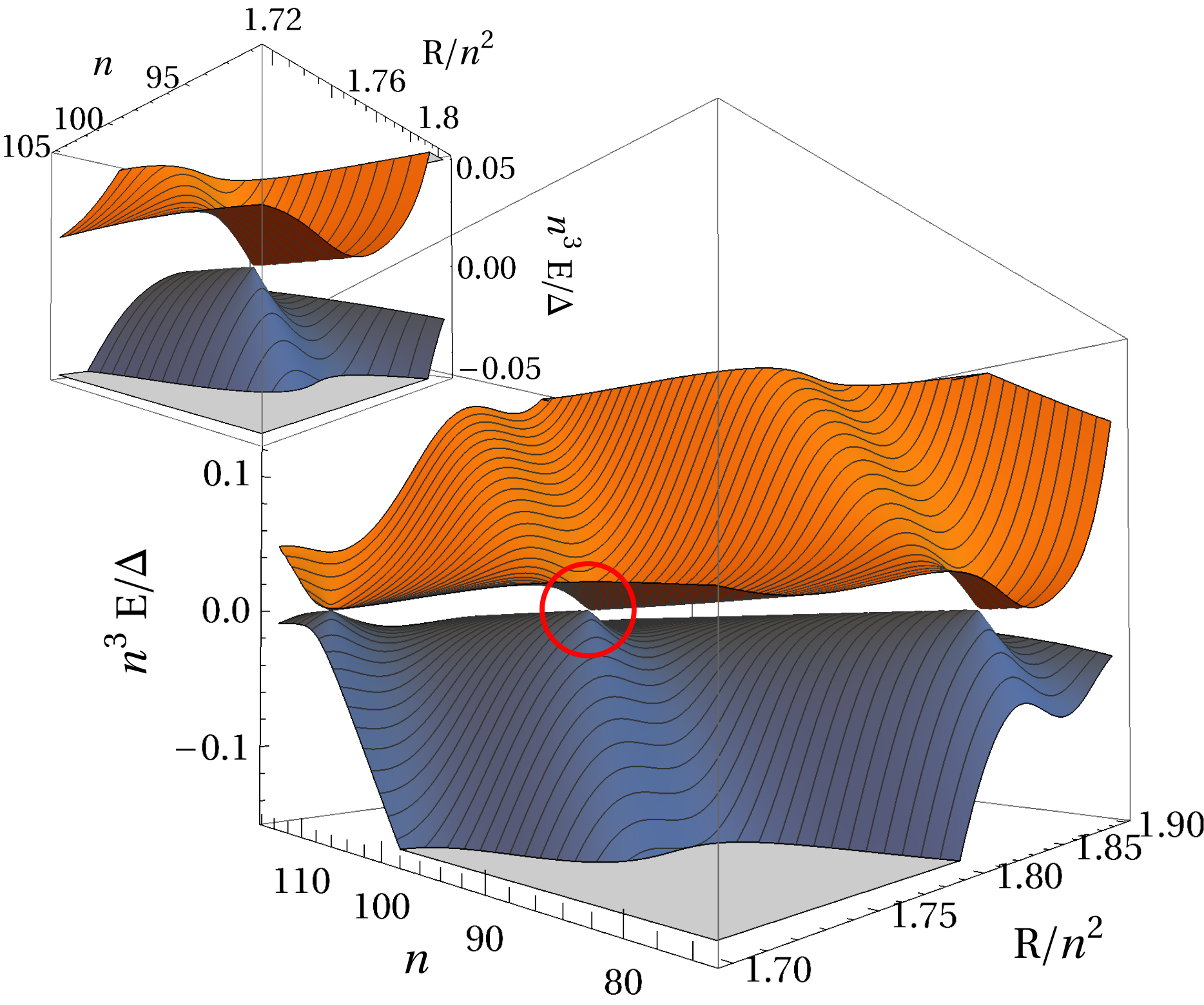}
	\caption{The adiabatic potential energy surfaces, for which the principle quantum number $n$ of the Rydberg atom serves as a synthetic parameter and dimension and the internuclear distance $R$ as the conventional parameter. 
	Conical intersections occur at $n_1=79$, $n_2=99$, and $n_3=111$, where the potentials become degenerate. Each mesh line corresponds to an integer $n$. The inset shows a zoom of the region in the vicinity of $n_2$.
	\label{fig:conic}}
\end{figure}

We search for CI in the PES, functions of both the real adiabatic parameter $R$ and the synthetic parameter $n$, by analytically summing $t(R)$ \cite{Chibisov2002,Eiles2019}, and then allowing $n$ to become a continuous variable by replacing the hydrogenic radial wave functions with Whittaker Coulomb functions valid for non-integer $n$. 
Figure \ref{fig:conic} shows these potential surfaces over a range of $n$.
In this region, the potentials are smoothly oscillatory functions of $R$ and $n$, but CI appear near the $n$ values $n_1=79$, $n_2=99$, and $n_3=111$. 
Here, we must include the effects of non-adiabatic coupling and interactions. 

In a beyond Born-Oppenheimer treatment, the non-adiabatic effects are manifested in the vibrational Schrö\-ding\-er equation as derivative coupling elements of first and second degree. 
For the electronic two-state problem, an analytical unitary transformation $U(n)$ exists such that the derivative terms are eliminated. 
$U$ represents a rotation about the angle
$\varphi(n,R) = \int_R^\infty \mathrm{d}R'\,\mathcal{P}(n,R')$, where the non-adiabatic derivative coupling matrix element $\mathcal{P}(n,R) = \braket{\psi_+|\nabla_R\psi_-}$ quantifies the coupling between the adiabatic states. 
$\nabla_R$ represents the gradient operator with respect to $R$.
Applying this transformation results in a non-diagonal potential in the \emph{diabatic representation} \cite{Koppel2007},
\begin{equation}
    V_\text{diabatic}(n,R) = \begin{pmatrix}q & \sqrt{q\cdot t} \\ \sqrt{q\cdot t} & t+\Delta
    \end{pmatrix}. \label{eq:H}
\end{equation}
The diabatic PEC $q$ and $t+\Delta$ are shown as dashed curves in Fig.~\ref{fig:scheme}a. 
The position $R_0$ of the avoided crossing of the adiabatic potentials is characterized by a local maximum of $\mathcal{P}$, which is illustrated in figure \ref{fig:scheme}c.
It typically coincides with a crossing of the diabatic PEC, which occurs at a position $R_0$  obtained from $d(n,R_0)=2\sqrt{q\cdot t}$, where $d(n,R)=|V_+-V_-|$ is the energy gap between the adiabatic potentials.
From the diabatic representation, the positions of CI can be obtained easily as the $(n,R)$ values where the off-diagonal coupling vanishes and the diagonal elements become degenerate. 
This is possible due to the oscillatory character of the potentials, since $q$ vanishes at a node of the $\ket{nl_0}$ wave function.
If, simultaneously, $t = -\Delta$, then the gap vanishes
$\lim_{R_0\rightarrow R_i} d(n_i,R)=0$,
and the derivative coupling diverges
$\lim_{R_0\rightarrow R_i} \mathcal{P}(n_i,R)=\infty.$
This is characteristic for a CI \cite{Worth2004}. Here, the Born-Oppenheimer approximation clearly breaks down. 
By aid of the principle quantum number $n$, we can tune $q$ and $t$ such that a CI can be realized.

We account for the probability of a non-adiabatic transition between the adiabatic potential surfaces at the avoided crossing via the semi-classical hopping probability \cite{Clark1979},
\begin{equation}
    P(n,v) = \exp\left(\frac{-\pi d(n,R_0)}{4v|\mathcal{P}(n,R_0)|}\right).
\end{equation}
The $l$-changing rate can now be extended to account for this non-adiabatic process, 
\begin{equation}
    \gamma(n,v)=\pi R_0^2 \rho v(1-P(n,v)).
\end{equation}
At the CI, $P(n_i,v) = 1$ and the analogue of radiationless transitions in polyatomic molecular systems occurs. 
Here, the Rydberg atom becomes \emph{transparent} to the colliding ground-state atom. 
\begin{figure}
	\centering
	\includegraphics[width=0.48\textwidth]{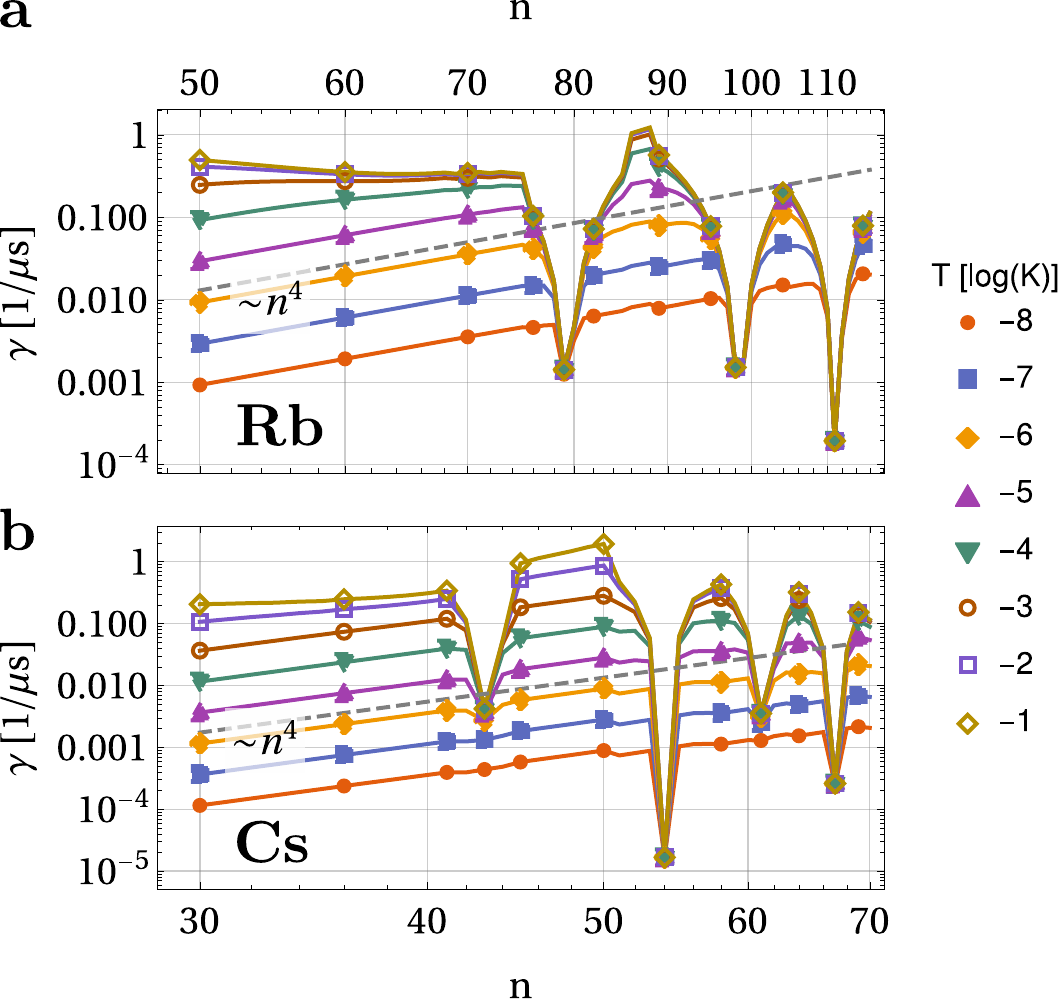}
		\caption{The rate of $l$-changing collisions in a (a) rubidium or (b) cesium gas, each with density $\rho=1.95\times 10^{12}\,\mathrm{cm}^{-3}$ for different temperatures from 10 nK (filled circles) to 0.1 K (diamonds). 
		The dashed, gray line shows the asymptotic behavior of the rates, proportional to $n^4$.
		At high temperatures, the rates become approximately independent of $n$. Very rapid variations in the rate as a function of $n$ occur in rubidium (a) at $n_1=79$, $n_2=99$, and $n_3=111$, where the rate decreases drastically and becomes independent of temperature and in cesium (b) at $n_1=43$, $n_2=54$, $n_3=61$, and $n_4=67$. Here, the decrease of the rate does not affect the lowest temperatures, in the case of $n_1$ and $n_3$.
		\label{fig:rate}}
\end{figure}

Figure \ref{fig:rate} shows the rate of $l$-changing collisions as a function of $n$ and across many decades of temperatures for both rubidium (a) and cesium (b) gases.
The density, $\rho=1.95\times10^{12}\,\mathrm{cm}^{-3}$, is typical for ultracold atomic clouds. 
The temperature determines the mean relative velocity of two atoms of the ensemble via $v=\sqrt{6k_BT/M}$, where $k_B$ is the Boltzmann constant and $M$ is the atomic mass. 
We focus first on the regime where no CI are found in the PES, namely $n<75$ in Rb and $n<40$ in Cs. 
Here, the splitting of the adiabatic functions scales as the difference in energy between the potential $V_-$ and the detuning $\Delta$, satisfying $d(n,R_0)\sim n^{-4}$.
Since no CI are present to induce rapid variation in the derivative coupling, it remains approximately constant as a function of $n$, $\mathcal{P}(n,R_0)\sim p$ (compare Fig.~\ref{fig:scheme}c). 
For low temperatures, the hopping probability is heavily suppressed,  $P(n,v)\to 0$, since $(vn^4p)^{-1}$ is large. 
In this limit, we obtain the adiabatic rate, proportional to the geometric cross section of the Rydberg atom $\sim n^4$, while at high temperatures, $(vn^4p)^{-1}$ is small. 
Taking the first order expression for $P(n,v)$ in this limit, we find that the rate becomes independent of $n$ and $v$. 
These two limiting cases are supported by the results shown in Fig.~\ref{fig:rate}.
We note that in this high temperature limit, our approach fails to account for many other physical processes which contribute to $l$-mixing and are not included in our two-level approximation, such as the effect of the ion--atom polarization potential, molecular ion formation, and $n$-changing collisions.
We present results for these high temperatures merely as a consistency check of the model. 
More comprehensive studies applicable to this thermal limit can be found in Refs.  \cite{Omont1977,Hickman1979,Fabrikant1986,Lebedev1996}. 
%We present these results merely as a consistency check. I DO NOT LIKE THIS PARAGRAPH NOW BUT SOMETHING LIKE IT SHOULD GO HERE

At larger $n$ values, this same behavior continues except for the notable exceptions around specific $n$ values, where the rate drops drastically and becomes nearly independent of the temperature. 
These are the points of collisional transparency induced by the synthetic conical intersections, where $l$-changing processes are strongly suppressed. 
Measurements of the $l$-changing collision rate are routinely taken, typically via field ionization schemes, and should be capable of directly probing these features. 

The points of collisional transparency are relevant for the preparation of circular Rydberg states \cite{Holzl2019}.
Away from the CI and at typical $\mu$K ultracold temperatures, the rapid rate of $l$-changing collisions, scaling as $n^4$, already exceeds typical radiative decay rates of Rydberg atoms, scaling as $n^3$. 
Thus, experimental schemes attempting to build applications on these high-$l$ Rydberg states, must either be performed at low densities or very low temperatures to avoid rapid decay of the $l_0$ state into higher-$l$ components. 
Apart from their fundamental interest, this gives the CI-induced collisional transparencies a crucial application since they suppress this rate over a wide range of temperatures and densities.

\begin{table}[b]
\setlength{\tabcolsep}{6pt}
\renewcommand{\arraystretch}{1.2}
    \centering
    \begin{tabular}{|c||c|c|c|c||c|}
        \hline
        species & $l_0$ & $\mu_{l_0}$ & $a(0)$ & $\alpha$ & $n_i$ \\ 
        \hline
        Li & 2 & 0.002129  & -6.7  & 164.2 & $\sim$ 200 \\
        Na & 2 & 0.015543  & -5.9  & 162.7 & 31, 38, 44 \\
        K  & 3 & 0.010098  & -15   & 290.6 & 116, 144 \\
        Rb & 3 & 0.0165192 & -15.2 & 319.2 & 79, 99, 111 \\
        Cs & 0 & 4.049425  & -21.8 & 401   & 43, 54, 61, 67 \\
        Cs & 3 & 0.033392  & -21.8 & 401   & 65, 75, 85 \\
        \hline
    \end{tabular}
    \caption{Table of alkaline atomic species and the occurrence of collisional transparency at the lowest few principle quantum numbers $n_i$ depending on the quantum defect $\mu_{l_0}$ \cite{Lorenzen1983,Han2006,Goy1982,Weber1987}, the zero-energy scattering length $a(0)$ \cite{Eiles2018,Karule1965,Engel2019,Sassmannshausen2015}, and the atomic polarizability $\alpha$ \cite{Miffre2006,Holmgren2010,Amini2003,Mitroy2010}.}
    \label{tab:species}
\end{table}

In table \ref{tab:species}, we present estimates for the occurrence of synthetic CI in the alkaline atomic species.
In each species, a low-$l$ state with angular momentum $l_0$ is selected such that its quantum defect $\mu_{l_0}$ comes closest to satisfying $\delta_s/\pi$. 
Note that two such states can satisfy this condition in Cs. 
Depending furthermore on the zero-energy scattering length $a(0)$ and the polarizability $\alpha$, we can predict the principle quantum numbers $n_i$ at which the corresponding collisional transparency is achieved. 
At large $n>70$, the effect of collisional transparency tends to be more pronounced in the sense that sub-$\mu$K temperatures are affected. 
As the electronic density of states increases with $n$, it is more likely that the synthetic CI, occurring in continuous $n$ space, occur at nearly integer values of $n$. 
At low $n$, this is a matter of chance. 
However, by selecting appropriate fine and hyperfine structure states (which are not accounted for in our analysis), one may further tune the quantum defect to achieve a stronger effect. 

In conclusion, we investigate low-temperature $l$-changing collisions of ground-state atoms with Rydberg atoms based on the analysis of the adiabatic Born-Oppenheimer PES underlying this process. 
These collisions modify the lifetime of Rydberg states, and along with the associated cross sections, provide avenues to probe fundamental atomic properties \cite{Niederprum2015,Schlagmuller2016x,Kanungo2020,Geppert2020}.
Emerging from the oscillatory character of both these PES and the corresponding non-adiabatic coupling elements, we find that certain principle quantum numbers $n_i$ exist where the $l$-changing rate is strongly suppressed, independently of temperature, leading to collisional transparency. 
This process can be explained by the existence of hidden conical intersections occurring in the vibrational coordinate $R$ and the synthetic dimension $n$.
This unconventional mechanism for the occurrence of CI in a diatomic setup provides a unique environment for tests of fundamental quantum processes such as vibronic couplings, radiationless decay, and geometric phases, with applications in Rydberg quantum chemistry.
Furthermore, our results provide immediate opportunities for the control of lifetimes of low-$l$ Rydberg states as well as means to circumvent unintended limitations in the preparation of these states and of circular Rydberg states. 
The collisional transparency allows also for precision measurements of quantum defects and low-energy $s$-wave scattering lengths of the underlying electron--atom interaction, because the positions of conical intersections at specific $n_i$ depend sensitively on these parameters.

\begin{acknowledgements}
    The authors gratefully acknowledge valuable discussions about the applicability of our results in the scope of high-$l$ state production with Tilman Pfau, Florian Meinert, Thomas Dieterle, and Christian Hölzl.
	ME acknowledges partial support from the Alexander von Humboldt Stiftung. 
	PS acknowledges support from the German Research Foundation (DFG) within the priority program "Giant Interactions in Rydberg Systems" [DFG SPP 1929 GiRyd project SCHM 885/30-1].
\end{acknowledgements}

\bibliographystyle{apsrev4-2}
%\bibliography{ulrm.bib}
\input{scit.bbl}

\end{document}

%% file: scit.bbl
%apsrev4-2.bst 2019-01-14 (MD) hand-edited version of apsrev4-1.bst
%Control: key (0)
%Control: author (72) initials jnrlst
%Control: editor formatted (1) identically to author
%Control: production of article title (-1) disabled
%Control: page (0) single
%Control: year (1) truncated
%Control: production of eprint (0) enabled
%

%% file: scit.bbl
\begin{thebibliography}{56}%
\makeatletter
\providecommand \@ifxundefined [1]{%
 \@ifx{#1\undefined}
}%
\providecommand \@ifnum [1]{%
 \ifnum #1\expandafter \@firstoftwo
 \else \expandafter \@secondoftwo
 \fi
}%
\providecommand \@ifx [1]{%
 \ifx #1\expandafter \@firstoftwo
 \else \expandafter \@secondoftwo
 \fi
}%
\providecommand \natexlab [1]{#1}%
\providecommand \enquote  [1]{``#1''}%
\providecommand \bibnamefont  [1]{#1}%
\providecommand \bibfnamefont [1]{#1}%
\providecommand \citenamefont [1]{#1}%
\providecommand \href@noop [0]{\@secondoftwo}%
\providecommand \href [0]{\begingroup \@sanitize@url \@href}%
\providecommand \@href[1]{\@@startlink{#1}\@@href}%
\providecommand \@@href[1]{\endgroup#1\@@endlink}%
\providecommand \@sanitize@url [0]{\catcode `\\12\catcode `\$12\catcode
  `\&12\catcode `\#12\catcode `\^12\catcode `\_12\catcode `\%12\relax}%
\providecommand \@@startlink[1]{}%
\providecommand \@@endlink[0]{}%
\providecommand \url  [0]{\begingroup\@sanitize@url \@url }%
\providecommand \@url [1]{\endgroup\@href {#1}{\urlprefix }}%
\providecommand \urlprefix  [0]{URL }%
\providecommand \Eprint [0]{\href }%
\providecommand \doibase [0]{https://doi.org/}%
\providecommand \selectlanguage [0]{\@gobble}%
\providecommand \bibinfo  [0]{\@secondoftwo}%
\providecommand \bibfield  [0]{\@secondoftwo}%
\providecommand \translation [1]{[#1]}%
\providecommand \BibitemOpen [0]{}%
\providecommand \bibitemStop [0]{}%
\providecommand \bibitemNoStop [0]{.\EOS\space}%
\providecommand \EOS [0]{\spacefactor3000\relax}%
\providecommand \BibitemShut  [1]{\csname bibitem#1\endcsname}%
\let\auto@bib@innerbib\@empty
%</preamble>
\bibitem [{\citenamefont {Born}\ and\ \citenamefont
  {Oppenheimer}(1927)}]{Born1927}%
  \BibitemOpen
  \bibfield  {author} {\bibinfo {author} {\bibfnamefont {M.}~\bibnamefont
  {Born}}\ and\ \bibinfo {author} {\bibfnamefont {R.}~\bibnamefont
  {Oppenheimer}},\ }\href
  {https://doi.org/https://doi.org/10.1002/andp.19273892002} {\bibfield
  {journal} {\bibinfo  {journal} {Ann. Phys.}\ }\textbf {\bibinfo {volume}
  {389}},\ \bibinfo {pages} {457} (\bibinfo {year} {1927})}\BibitemShut
  {NoStop}%
\bibitem [{\citenamefont {Köppel}\ \emph {et~al.}(1984)\citenamefont
  {Köppel}, \citenamefont {Domcke},\ and\ \citenamefont
  {Cederbaum}}]{Koppel2007}%
  \BibitemOpen
  \bibfield  {author} {\bibinfo {author} {\bibfnamefont {H.}~\bibnamefont
  {Köppel}}, \bibinfo {author} {\bibfnamefont {W.}~\bibnamefont {Domcke}},\
  and\ \bibinfo {author} {\bibfnamefont {L.~S.}\ \bibnamefont {Cederbaum}},\
  }\bibinfo {title} {Multimode molecular dynamics beyond the {Born-Oppenheimer}
  approximation},\ in\ \href {https://doi.org/10.1002/9780470142813.ch2} {\emph
  {\bibinfo {booktitle} {Advances in Chemical Physics}}}\ (\bibinfo
  {publisher} {John Wiley \& Sons, Ltd},\ \bibinfo {year} {1984})\ pp.\
  \bibinfo {pages} {59--246}\BibitemShut {NoStop}%
\bibitem [{\citenamefont {Worth}\ and\ \citenamefont
  {Cederbaum}(2004)}]{Worth2004}%
  \BibitemOpen
  \bibfield  {author} {\bibinfo {author} {\bibfnamefont {G.~A.}\ \bibnamefont
  {Worth}}\ and\ \bibinfo {author} {\bibfnamefont {L.~S.}\ \bibnamefont
  {Cederbaum}},\ }\href
  {https://doi.org/10.1146/annurev.physchem.55.091602.094335} {\bibfield
  {journal} {\bibinfo  {journal} {Annu. Rev. Phys. Chem.}\ }\textbf {\bibinfo
  {volume} {55}},\ \bibinfo {pages} {127} (\bibinfo {year} {2004})}\BibitemShut
  {NoStop}%
\bibitem [{\citenamefont {Worth}\ \emph {et~al.}(2008)\citenamefont {Worth},
  \citenamefont {Meyer}, \citenamefont {Köppel}, \citenamefont {Cederbaum},\
  and\ \citenamefont {Burghardt}}]{Worth2008}%
  \BibitemOpen
  \bibfield  {author} {\bibinfo {author} {\bibfnamefont {G.~A.}\ \bibnamefont
  {Worth}}, \bibinfo {author} {\bibfnamefont {H.-D.}\ \bibnamefont {Meyer}},
  \bibinfo {author} {\bibfnamefont {H.}~\bibnamefont {Köppel}}, \bibinfo
  {author} {\bibfnamefont {L.~S.}\ \bibnamefont {Cederbaum}},\ and\ \bibinfo
  {author} {\bibfnamefont {I.}~\bibnamefont {Burghardt}},\ }\href
  {https://doi.org/10.1080/01442350802137656} {\bibfield  {journal} {\bibinfo
  {journal} {International Reviews in Physical Chemistry}\ }\textbf {\bibinfo
  {volume} {27}},\ \bibinfo {pages} {569} (\bibinfo {year} {2008})}\BibitemShut
  {NoStop}%
\bibitem [{\citenamefont {Barbatti}\ \emph {et~al.}(2010)\citenamefont
  {Barbatti}, \citenamefont {Aquino}, \citenamefont {Szymczak}, \citenamefont
  {Nachtigallov{\'a}}, \citenamefont {Hobza},\ and\ \citenamefont
  {Lischka}}]{Barbatti2010}%
  \BibitemOpen
  \bibfield  {author} {\bibinfo {author} {\bibfnamefont {M.}~\bibnamefont
  {Barbatti}}, \bibinfo {author} {\bibfnamefont {A.~J.~A.}\ \bibnamefont
  {Aquino}}, \bibinfo {author} {\bibfnamefont {J.~J.}\ \bibnamefont
  {Szymczak}}, \bibinfo {author} {\bibfnamefont {D.}~\bibnamefont
  {Nachtigallov{\'a}}}, \bibinfo {author} {\bibfnamefont {P.}~\bibnamefont
  {Hobza}},\ and\ \bibinfo {author} {\bibfnamefont {H.}~\bibnamefont
  {Lischka}},\ }\href {https://doi.org/10.1073/pnas.1014982107} {\bibfield
  {journal} {\bibinfo  {journal} {Proc. Nat. Acad. Sci.}\ }\textbf {\bibinfo
  {volume} {107}},\ \bibinfo {pages} {21453} (\bibinfo {year}
  {2010})}\BibitemShut {NoStop}%
\bibitem [{\citenamefont {Hammarström}\ and\ \citenamefont
  {Styring}(2008)}]{Hammarstrom2008}%
  \BibitemOpen
  \bibfield  {author} {\bibinfo {author} {\bibfnamefont {L.}~\bibnamefont
  {Hammarström}}\ and\ \bibinfo {author} {\bibfnamefont {S.}~\bibnamefont
  {Styring}},\ }\href {https://doi.org/10.1098/rstb.2007.2225} {\bibfield
  {journal} {\bibinfo  {journal} {Phil. Trans. R. Soc. B}\ }\textbf {\bibinfo
  {volume} {363}},\ \bibinfo {pages} {1283} (\bibinfo {year}
  {2008})}\BibitemShut {NoStop}%
\bibitem [{\citenamefont {Arnold}\ \emph {et~al.}(2018)\citenamefont {Arnold},
  \citenamefont {Vendrell}, \citenamefont {Welsch},\ and\ \citenamefont
  {Santra}}]{Arnold2018}%
  \BibitemOpen
  \bibfield  {author} {\bibinfo {author} {\bibfnamefont {C.}~\bibnamefont
  {Arnold}}, \bibinfo {author} {\bibfnamefont {O.}~\bibnamefont {Vendrell}},
  \bibinfo {author} {\bibfnamefont {R.}~\bibnamefont {Welsch}},\ and\ \bibinfo
  {author} {\bibfnamefont {R.}~\bibnamefont {Santra}},\ }\href
  {https://doi.org/10.1103/PhysRevLett.120.123001} {\bibfield  {journal}
  {\bibinfo  {journal} {Phys. Rev. Lett.}\ }\textbf {\bibinfo {volume} {120}},\
  \bibinfo {pages} {123001} (\bibinfo {year} {2018})}\BibitemShut {NoStop}%
\bibitem [{\citenamefont {Mabrouk}\ \emph {et~al.}(2020)\citenamefont
  {Mabrouk}, \citenamefont {Zrafi},\ and\ \citenamefont
  {Berriche}}]{Mabrouk2020}%
  \BibitemOpen
  \bibfield  {author} {\bibinfo {author} {\bibfnamefont {N.}~\bibnamefont
  {Mabrouk}}, \bibinfo {author} {\bibfnamefont {W.}~\bibnamefont {Zrafi}},\
  and\ \bibinfo {author} {\bibfnamefont {H.}~\bibnamefont {Berriche}},\ }\href
  {https://doi.org/10.1080/00268976.2019.1605098} {\bibfield  {journal}
  {\bibinfo  {journal} {Mol. Phys.}\ }\textbf {\bibinfo {volume} {118}},\
  \bibinfo {pages} {e1605098} (\bibinfo {year} {2020})}\BibitemShut {NoStop}%
\bibitem [{\citenamefont {Moiseyev}\ \emph {et~al.}(2008)\citenamefont
  {Moiseyev}, \citenamefont {{\v{S}}indelka},\ and\ \citenamefont
  {Cederbaum}}]{Moiseyev2008}%
  \BibitemOpen
  \bibfield  {author} {\bibinfo {author} {\bibfnamefont {N.}~\bibnamefont
  {Moiseyev}}, \bibinfo {author} {\bibfnamefont {M.}~\bibnamefont
  {{\v{S}}indelka}},\ and\ \bibinfo {author} {\bibfnamefont {L.~S.}\
  \bibnamefont {Cederbaum}},\ }\href
  {https://doi.org/10.1088/0953-4075/41/22/221001} {\bibfield  {journal}
  {\bibinfo  {journal} {J. Phys. B: At. Mol. Opt. Phys.}\ }\textbf {\bibinfo
  {volume} {41}},\ \bibinfo {pages} {221001} (\bibinfo {year}
  {2008})}\BibitemShut {NoStop}%
\bibitem [{\citenamefont {{\v{S}}indelka}\ \emph {et~al.}(2011)\citenamefont
  {{\v{S}}indelka}, \citenamefont {Moiseyev},\ and\ \citenamefont
  {Cederbaum}}]{Sindelka2011}%
  \BibitemOpen
  \bibfield  {author} {\bibinfo {author} {\bibfnamefont {M.}~\bibnamefont
  {{\v{S}}indelka}}, \bibinfo {author} {\bibfnamefont {N.}~\bibnamefont
  {Moiseyev}},\ and\ \bibinfo {author} {\bibfnamefont {L.~S.}\ \bibnamefont
  {Cederbaum}},\ }\href {https://doi.org/10.1088/0953-4075/44/4/045603}
  {\bibfield  {journal} {\bibinfo  {journal} {J. Phys. B: At. Mol. Opt. Phys.}\
  }\textbf {\bibinfo {volume} {44}},\ \bibinfo {pages} {045603} (\bibinfo
  {year} {2011})}\BibitemShut {NoStop}%
\bibitem [{\citenamefont {W\"uster}\ \emph {et~al.}(2011)\citenamefont
  {W\"uster}, \citenamefont {Eisfeld},\ and\ \citenamefont
  {Rost}}]{Wuster2011}%
  \BibitemOpen
  \bibfield  {author} {\bibinfo {author} {\bibfnamefont {S.}~\bibnamefont
  {W\"uster}}, \bibinfo {author} {\bibfnamefont {A.}~\bibnamefont {Eisfeld}},\
  and\ \bibinfo {author} {\bibfnamefont {J.~M.}\ \bibnamefont {Rost}},\ }\href
  {https://doi.org/10.1103/PhysRevLett.106.153002} {\bibfield  {journal}
  {\bibinfo  {journal} {Phys. Rev. Lett.}\ }\textbf {\bibinfo {volume} {106}},\
  \bibinfo {pages} {153002} (\bibinfo {year} {2011})}\BibitemShut {NoStop}%
\bibitem [{\citenamefont {Gambetta}\ \emph {et~al.}(2020)\citenamefont
  {Gambetta}, \citenamefont {Zhang}, \citenamefont {Hennrich}, \citenamefont
  {Lesanovsky},\ and\ \citenamefont {Li}}]{Gambetta2020}%
  \BibitemOpen
  \bibfield  {author} {\bibinfo {author} {\bibfnamefont {F.~M.}\ \bibnamefont
  {Gambetta}}, \bibinfo {author} {\bibfnamefont {C.}~\bibnamefont {Zhang}},
  \bibinfo {author} {\bibfnamefont {M.}~\bibnamefont {Hennrich}}, \bibinfo
  {author} {\bibfnamefont {I.}~\bibnamefont {Lesanovsky}},\ and\ \bibinfo
  {author} {\bibfnamefont {W.}~\bibnamefont {Li}},\ }\href@noop {} {\bibinfo
  {title} {Exploring the many-body dynamics near a conical intersection with
  trapped rydberg ions}} (\bibinfo {year} {2020}),\ \Eprint
  {https://arxiv.org/abs/2012.01834} {arXiv:2012.01834 [physics.atom-ph]}
  \BibitemShut {NoStop}%
\bibitem [{\citenamefont {von Neumann}\ and\ \citenamefont
  {Wigner}(1993)}]{vonNeumann1993}%
  \BibitemOpen
  \bibfield  {author} {\bibinfo {author} {\bibfnamefont {J.}~\bibnamefont {von
  Neumann}}\ and\ \bibinfo {author} {\bibfnamefont {E.~P.}\ \bibnamefont
  {Wigner}},\ }\bibinfo {title} {Über merkwürdige diskrete eigenwerte},\ in\
  \href {https://doi.org/10.1007/978-3-662-02781-3_19} {\emph {\bibinfo
  {booktitle} {The Collected Works of Eugene Paul Wigner: Part A: The
  Scientific Papers}}},\ \bibinfo {editor} {edited by\ \bibinfo {editor}
  {\bibfnamefont {A.~S.}\ \bibnamefont {Wightman}}}\ (\bibinfo  {publisher}
  {Springer Berlin Heidelberg},\ \bibinfo {address} {Berlin, Heidelberg},\
  \bibinfo {year} {1993})\ pp.\ \bibinfo {pages} {291--293}\BibitemShut
  {NoStop}%
\bibitem [{\citenamefont {Fermi}(1934)}]{Fermi1934}%
  \BibitemOpen
  \bibfield  {author} {\bibinfo {author} {\bibfnamefont {E.}~\bibnamefont
  {Fermi}},\ }\href {https://doi.org/10.1007/BF02959829} {\bibfield  {journal}
  {\bibinfo  {journal} {Nuovo Cim.}\ }\textbf {\bibinfo {volume} {11}},\
  \bibinfo {pages} {157} (\bibinfo {year} {1934})}\BibitemShut {NoStop}%
\bibitem [{\citenamefont {Omont}(1977)}]{Omont1977}%
  \BibitemOpen
  \bibfield  {author} {\bibinfo {author} {\bibfnamefont {A.}~\bibnamefont
  {Omont}},\ }\href {https://doi.org/10.1051/jphys:0197700380110134300}
  {\bibfield  {journal} {\bibinfo  {journal} {J. Phys. France}\ }\textbf
  {\bibinfo {volume} {38}},\ \bibinfo {pages} {1343} (\bibinfo {year}
  {1977})}\BibitemShut {NoStop}%
\bibitem [{\citenamefont {Hickman}(1979)}]{Hickman1979}%
  \BibitemOpen
  \bibfield  {author} {\bibinfo {author} {\bibfnamefont {A.~P.}\ \bibnamefont
  {Hickman}},\ }\href {https://doi.org/10.1103/PhysRevA.19.994} {\bibfield
  {journal} {\bibinfo  {journal} {Phys. Rev. A}\ }\textbf {\bibinfo {volume}
  {19}},\ \bibinfo {pages} {994} (\bibinfo {year} {1979})}\BibitemShut
  {NoStop}%
\bibitem [{\citenamefont {Gallagher}(1988)}]{Gallagher1988}%
  \BibitemOpen
  \bibfield  {author} {\bibinfo {author} {\bibfnamefont {T.~F.}\ \bibnamefont
  {Gallagher}},\ }\href {https://doi.org/10.1088/0034-4885/51/2/001} {\bibfield
   {journal} {\bibinfo  {journal} {Rep. Prog. Phys.}\ }\textbf {\bibinfo
  {volume} {51}},\ \bibinfo {pages} {143} (\bibinfo {year} {1988})}\BibitemShut
  {NoStop}%
\bibitem [{\citenamefont {Lebedev}\ and\ \citenamefont
  {Fabrikant}(1996)}]{Lebedev1996}%
  \BibitemOpen
  \bibfield  {author} {\bibinfo {author} {\bibfnamefont {V.~S.}\ \bibnamefont
  {Lebedev}}\ and\ \bibinfo {author} {\bibfnamefont {I.~I.}\ \bibnamefont
  {Fabrikant}},\ }\href {https://doi.org/10.1103/PhysRevA.54.2888} {\bibfield
  {journal} {\bibinfo  {journal} {Phys. Rev. A}\ }\textbf {\bibinfo {volume}
  {54}},\ \bibinfo {pages} {2888} (\bibinfo {year} {1996})}\BibitemShut
  {NoStop}%
\bibitem [{\citenamefont {Greene}\ \emph {et~al.}(2000)\citenamefont {Greene},
  \citenamefont {Dickinson},\ and\ \citenamefont {Sadeghpour}}]{Greene2000}%
  \BibitemOpen
  \bibfield  {author} {\bibinfo {author} {\bibfnamefont {C.~H.}\ \bibnamefont
  {Greene}}, \bibinfo {author} {\bibfnamefont {A.~S.}\ \bibnamefont
  {Dickinson}},\ and\ \bibinfo {author} {\bibfnamefont {H.~R.}\ \bibnamefont
  {Sadeghpour}},\ }\href {https://doi.org/10.1103/PhysRevLett.85.2458}
  {\bibfield  {journal} {\bibinfo  {journal} {Phys. Rev. Lett.}\ }\textbf
  {\bibinfo {volume} {85}},\ \bibinfo {pages} {2458} (\bibinfo {year}
  {2000})}\BibitemShut {NoStop}%
\bibitem [{\citenamefont {Bendkowsky}\ \emph {et~al.}(2009)\citenamefont
  {Bendkowsky}, \citenamefont {Butscher}, \citenamefont {Nipper}, \citenamefont
  {Shaffer}, \citenamefont {Löw},\ and\ \citenamefont
  {Pfau}}]{Bendkowsky2009}%
  \BibitemOpen
  \bibfield  {author} {\bibinfo {author} {\bibfnamefont {V.}~\bibnamefont
  {Bendkowsky}}, \bibinfo {author} {\bibfnamefont {B.}~\bibnamefont
  {Butscher}}, \bibinfo {author} {\bibfnamefont {J.}~\bibnamefont {Nipper}},
  \bibinfo {author} {\bibfnamefont {J.~P.}\ \bibnamefont {Shaffer}}, \bibinfo
  {author} {\bibfnamefont {R.}~\bibnamefont {Löw}},\ and\ \bibinfo {author}
  {\bibfnamefont {T.}~\bibnamefont {Pfau}},\ }\href
  {https://doi.org/10.1038/nature07945} {\bibfield  {journal} {\bibinfo
  {journal} {Nature}\ }\textbf {\bibinfo {volume} {458}},\ \bibinfo {pages}
  {1005} (\bibinfo {year} {2009})}\BibitemShut {NoStop}%
\bibitem [{\citenamefont {Fey}\ \emph {et~al.}(2020)\citenamefont {Fey},
  \citenamefont {Hummel},\ and\ \citenamefont {Schmelcher}}]{Fey2019rev}%
  \BibitemOpen
  \bibfield  {author} {\bibinfo {author} {\bibfnamefont {C.}~\bibnamefont
  {Fey}}, \bibinfo {author} {\bibfnamefont {F.}~\bibnamefont {Hummel}},\ and\
  \bibinfo {author} {\bibfnamefont {P.}~\bibnamefont {Schmelcher}},\ }\href
  {https://doi.org/10.1080/00268976.2019.1679401} {\bibfield  {journal}
  {\bibinfo  {journal} {Mol. Phys.}\ }\textbf {\bibinfo {volume} {118}},\
  \bibinfo {pages} {e1679401} (\bibinfo {year} {2020})}\BibitemShut {NoStop}%
\bibitem [{\citenamefont {Eiles}(2019)}]{Eiles2019}%
  \BibitemOpen
  \bibfield  {author} {\bibinfo {author} {\bibfnamefont {M.~T.}\ \bibnamefont
  {Eiles}},\ }\href {https://doi.org/10.1088/1361-6455/ab19ca} {\bibfield
  {journal} {\bibinfo  {journal} {J. Phys. B: At. Mol. Opt. Phys.}\ }\textbf
  {\bibinfo {volume} {52}},\ \bibinfo {pages} {113001} (\bibinfo {year}
  {2019})}\BibitemShut {NoStop}%
\bibitem [{\citenamefont {Levi}\ \emph {et~al.}(2011)\citenamefont {Levi},
  \citenamefont {Rechtsman}, \citenamefont {Freedman}, \citenamefont
  {Schwartz}, \citenamefont {Manela},\ and\ \citenamefont {Segev}}]{Levi2011}%
  \BibitemOpen
  \bibfield  {author} {\bibinfo {author} {\bibfnamefont {L.}~\bibnamefont
  {Levi}}, \bibinfo {author} {\bibfnamefont {M.}~\bibnamefont {Rechtsman}},
  \bibinfo {author} {\bibfnamefont {B.}~\bibnamefont {Freedman}}, \bibinfo
  {author} {\bibfnamefont {T.}~\bibnamefont {Schwartz}}, \bibinfo {author}
  {\bibfnamefont {O.}~\bibnamefont {Manela}},\ and\ \bibinfo {author}
  {\bibfnamefont {M.}~\bibnamefont {Segev}},\ }\href
  {https://doi.org/10.1126/science.1202977} {\bibfield  {journal} {\bibinfo
  {journal} {Science}\ }\textbf {\bibinfo {volume} {332}},\ \bibinfo {pages}
  {1541} (\bibinfo {year} {2011})}\BibitemShut {NoStop}%
\bibitem [{\citenamefont {Kolkowitz}\ \emph {et~al.}(2017)\citenamefont
  {Kolkowitz}, \citenamefont {Bromley}, \citenamefont {Bothwell}, \citenamefont
  {Wall}, \citenamefont {Marti}, \citenamefont {Koller}, \citenamefont {Zhang},
  \citenamefont {Rey},\ and\ \citenamefont {Ye}}]{Kolkowitz2017}%
  \BibitemOpen
  \bibfield  {author} {\bibinfo {author} {\bibfnamefont {S.}~\bibnamefont
  {Kolkowitz}}, \bibinfo {author} {\bibfnamefont {S.~L.}\ \bibnamefont
  {Bromley}}, \bibinfo {author} {\bibfnamefont {T.}~\bibnamefont {Bothwell}},
  \bibinfo {author} {\bibfnamefont {M.~L.}\ \bibnamefont {Wall}}, \bibinfo
  {author} {\bibfnamefont {G.~E.}\ \bibnamefont {Marti}}, \bibinfo {author}
  {\bibfnamefont {A.~P.}\ \bibnamefont {Koller}}, \bibinfo {author}
  {\bibfnamefont {X.}~\bibnamefont {Zhang}}, \bibinfo {author} {\bibfnamefont
  {M.}~\bibnamefont {Rey}},\ and\ \bibinfo {author} {\bibfnamefont
  {J.}~\bibnamefont {Ye}},\ }\href {https://doi.org/10.1038/nature20811}
  {\bibfield  {journal} {\bibinfo  {journal} {Nature}\ }\textbf {\bibinfo
  {volume} {542}},\ \bibinfo {pages} {66} (\bibinfo {year} {2017})}\BibitemShut
  {NoStop}%
\bibitem [{\citenamefont {Celi}\ \emph {et~al.}(2014)\citenamefont {Celi},
  \citenamefont {Massignan}, \citenamefont {Ruseckas}, \citenamefont {Goldman},
  \citenamefont {Spielman}, \citenamefont {Juzeli\={u}nas},\ and\ \citenamefont
  {Lewenstein}}]{Celi2014}%
  \BibitemOpen
  \bibfield  {author} {\bibinfo {author} {\bibfnamefont {A.}~\bibnamefont
  {Celi}}, \bibinfo {author} {\bibfnamefont {P.}~\bibnamefont {Massignan}},
  \bibinfo {author} {\bibfnamefont {J.}~\bibnamefont {Ruseckas}}, \bibinfo
  {author} {\bibfnamefont {N.}~\bibnamefont {Goldman}}, \bibinfo {author}
  {\bibfnamefont {I.~B.}\ \bibnamefont {Spielman}}, \bibinfo {author}
  {\bibfnamefont {G.}~\bibnamefont {Juzeli\={u}nas}},\ and\ \bibinfo {author}
  {\bibfnamefont {M.}~\bibnamefont {Lewenstein}},\ }\href
  {https://doi.org/10.1103/PhysRevLett.112.043001} {\bibfield  {journal}
  {\bibinfo  {journal} {Phys. Rev. Lett.}\ }\textbf {\bibinfo {volume} {112}},\
  \bibinfo {pages} {043001} (\bibinfo {year} {2014})}\BibitemShut {NoStop}%
\bibitem [{\citenamefont {Boada}\ \emph {et~al.}(2012)\citenamefont {Boada},
  \citenamefont {Celi}, \citenamefont {Latorre},\ and\ \citenamefont
  {Lewenstein}}]{Boada2012}%
  \BibitemOpen
  \bibfield  {author} {\bibinfo {author} {\bibfnamefont {O.}~\bibnamefont
  {Boada}}, \bibinfo {author} {\bibfnamefont {A.}~\bibnamefont {Celi}},
  \bibinfo {author} {\bibfnamefont {J.~I.}\ \bibnamefont {Latorre}},\ and\
  \bibinfo {author} {\bibfnamefont {M.}~\bibnamefont {Lewenstein}},\ }\href
  {https://doi.org/10.1103/PhysRevLett.108.133001} {\bibfield  {journal}
  {\bibinfo  {journal} {Phys. Rev. Lett.}\ }\textbf {\bibinfo {volume} {108}},\
  \bibinfo {pages} {133001} (\bibinfo {year} {2012})}\BibitemShut {NoStop}%
\bibitem [{\citenamefont {Ozawa}\ \emph {et~al.}(2019)\citenamefont {Ozawa},
  \citenamefont {Price}, \citenamefont {Amo}, \citenamefont {Goldman},
  \citenamefont {Hafezi}, \citenamefont {Lu}, \citenamefont {Rechtsman},
  \citenamefont {Schuster}, \citenamefont {Simon}, \citenamefont {Zilberberg},\
  and\ \citenamefont {Carusotto}}]{Ozawa2019}%
  \BibitemOpen
  \bibfield  {author} {\bibinfo {author} {\bibfnamefont {T.}~\bibnamefont
  {Ozawa}}, \bibinfo {author} {\bibfnamefont {H.~M.}\ \bibnamefont {Price}},
  \bibinfo {author} {\bibfnamefont {A.}~\bibnamefont {Amo}}, \bibinfo {author}
  {\bibfnamefont {N.}~\bibnamefont {Goldman}}, \bibinfo {author} {\bibfnamefont
  {M.}~\bibnamefont {Hafezi}}, \bibinfo {author} {\bibfnamefont
  {L.}~\bibnamefont {Lu}}, \bibinfo {author} {\bibfnamefont {M.~C.}\
  \bibnamefont {Rechtsman}}, \bibinfo {author} {\bibfnamefont {D.}~\bibnamefont
  {Schuster}}, \bibinfo {author} {\bibfnamefont {J.}~\bibnamefont {Simon}},
  \bibinfo {author} {\bibfnamefont {O.}~\bibnamefont {Zilberberg}},\ and\
  \bibinfo {author} {\bibfnamefont {I.}~\bibnamefont {Carusotto}},\ }\href
  {https://doi.org/10.1103/RevModPhys.91.015006} {\bibfield  {journal}
  {\bibinfo  {journal} {Rev. Mod. Phys.}\ }\textbf {\bibinfo {volume} {91}},\
  \bibinfo {pages} {015006} (\bibinfo {year} {2019})}\BibitemShut {NoStop}%
\bibitem [{\citenamefont {Anderson}\ \emph {et~al.}(2013)\citenamefont
  {Anderson}, \citenamefont {Schwarzkopf}, \citenamefont {Sapiro},\ and\
  \citenamefont {Raithel}}]{Anderson2013}%
  \BibitemOpen
  \bibfield  {author} {\bibinfo {author} {\bibfnamefont {D.~A.}\ \bibnamefont
  {Anderson}}, \bibinfo {author} {\bibfnamefont {A.}~\bibnamefont
  {Schwarzkopf}}, \bibinfo {author} {\bibfnamefont {R.~E.}\ \bibnamefont
  {Sapiro}},\ and\ \bibinfo {author} {\bibfnamefont {G.}~\bibnamefont
  {Raithel}},\ }\href {https://doi.org/10.1103/PhysRevA.88.031401} {\bibfield
  {journal} {\bibinfo  {journal} {Phys. Rev. A}\ }\textbf {\bibinfo {volume}
  {88}},\ \bibinfo {pages} {031401(R)} (\bibinfo {year} {2013})}\BibitemShut
  {NoStop}%
\bibitem [{\citenamefont {Nguyen}\ \emph {et~al.}(2018)\citenamefont {Nguyen},
  \citenamefont {Raimond}, \citenamefont {Sayrin}, \citenamefont {Corti\~nas},
  \citenamefont {Cantat-Moltrecht}, \citenamefont {Assemat}, \citenamefont
  {Dotsenko}, \citenamefont {Gleyzes}, \citenamefont {Haroche}, \citenamefont
  {Roux}, \citenamefont {Jolicoeur},\ and\ \citenamefont {Brune}}]{Nguyen2018}%
  \BibitemOpen
  \bibfield  {author} {\bibinfo {author} {\bibfnamefont {T.~L.}\ \bibnamefont
  {Nguyen}}, \bibinfo {author} {\bibfnamefont {J.~M.}\ \bibnamefont {Raimond}},
  \bibinfo {author} {\bibfnamefont {C.}~\bibnamefont {Sayrin}}, \bibinfo
  {author} {\bibfnamefont {R.}~\bibnamefont {Corti\~nas}}, \bibinfo {author}
  {\bibfnamefont {T.}~\bibnamefont {Cantat-Moltrecht}}, \bibinfo {author}
  {\bibfnamefont {F.}~\bibnamefont {Assemat}}, \bibinfo {author} {\bibfnamefont
  {I.}~\bibnamefont {Dotsenko}}, \bibinfo {author} {\bibfnamefont
  {S.}~\bibnamefont {Gleyzes}}, \bibinfo {author} {\bibfnamefont
  {S.}~\bibnamefont {Haroche}}, \bibinfo {author} {\bibfnamefont
  {G.}~\bibnamefont {Roux}}, \bibinfo {author} {\bibfnamefont {T.}~\bibnamefont
  {Jolicoeur}},\ and\ \bibinfo {author} {\bibfnamefont {M.}~\bibnamefont
  {Brune}},\ }\href {https://doi.org/10.1103/PhysRevX.8.011032} {\bibfield
  {journal} {\bibinfo  {journal} {Phys. Rev. X}\ }\textbf {\bibinfo {volume}
  {8}},\ \bibinfo {pages} {011032} (\bibinfo {year} {2018})}\BibitemShut
  {NoStop}%
\bibitem [{\citenamefont {Corti\~nas}\ \emph {et~al.}(2020)\citenamefont
  {Corti\~nas}, \citenamefont {Favier}, \citenamefont {Ravon}, \citenamefont
  {M\'ehaignerie}, \citenamefont {Machu}, \citenamefont {Raimond},
  \citenamefont {Sayrin},\ and\ \citenamefont {Brune}}]{Cortinas2020}%
  \BibitemOpen
  \bibfield  {author} {\bibinfo {author} {\bibfnamefont {R.~G.}\ \bibnamefont
  {Corti\~nas}}, \bibinfo {author} {\bibfnamefont {M.}~\bibnamefont {Favier}},
  \bibinfo {author} {\bibfnamefont {B.}~\bibnamefont {Ravon}}, \bibinfo
  {author} {\bibfnamefont {P.}~\bibnamefont {M\'ehaignerie}}, \bibinfo {author}
  {\bibfnamefont {Y.}~\bibnamefont {Machu}}, \bibinfo {author} {\bibfnamefont
  {J.~M.}\ \bibnamefont {Raimond}}, \bibinfo {author} {\bibfnamefont
  {C.}~\bibnamefont {Sayrin}},\ and\ \bibinfo {author} {\bibfnamefont
  {M.}~\bibnamefont {Brune}},\ }\href
  {https://doi.org/10.1103/PhysRevLett.124.123201} {\bibfield  {journal}
  {\bibinfo  {journal} {Phys. Rev. Lett.}\ }\textbf {\bibinfo {volume} {124}},\
  \bibinfo {pages} {123201} (\bibinfo {year} {2020})}\BibitemShut {NoStop}%
\bibitem [{\citenamefont {Cantat-Moltrecht}\ \emph {et~al.}(2020)\citenamefont
  {Cantat-Moltrecht}, \citenamefont {Corti\~nas}, \citenamefont {Ravon},
  \citenamefont {M\'ehaignerie}, \citenamefont {Haroche}, \citenamefont
  {Raimond}, \citenamefont {Favier}, \citenamefont {Brune},\ and\ \citenamefont
  {Sayrin}}]{Cantat2020}%
  \BibitemOpen
  \bibfield  {author} {\bibinfo {author} {\bibfnamefont {T.}~\bibnamefont
  {Cantat-Moltrecht}}, \bibinfo {author} {\bibfnamefont {R.}~\bibnamefont
  {Corti\~nas}}, \bibinfo {author} {\bibfnamefont {B.}~\bibnamefont {Ravon}},
  \bibinfo {author} {\bibfnamefont {P.}~\bibnamefont {M\'ehaignerie}}, \bibinfo
  {author} {\bibfnamefont {S.}~\bibnamefont {Haroche}}, \bibinfo {author}
  {\bibfnamefont {J.~M.}\ \bibnamefont {Raimond}}, \bibinfo {author}
  {\bibfnamefont {M.}~\bibnamefont {Favier}}, \bibinfo {author} {\bibfnamefont
  {M.}~\bibnamefont {Brune}},\ and\ \bibinfo {author} {\bibfnamefont
  {C.}~\bibnamefont {Sayrin}},\ }\href
  {https://doi.org/10.1103/PhysRevResearch.2.022032} {\bibfield  {journal}
  {\bibinfo  {journal} {Phys. Rev. Research}\ }\textbf {\bibinfo {volume}
  {2}},\ \bibinfo {pages} {022032(R)} (\bibinfo {year} {2020})}\BibitemShut
  {NoStop}%
\bibitem [{\citenamefont {Hamilton}\ \emph {et~al.}(2002)\citenamefont
  {Hamilton}, \citenamefont {Greene},\ and\ \citenamefont
  {Sadeghpour}}]{Hamilton2002}%
  \BibitemOpen
  \bibfield  {author} {\bibinfo {author} {\bibfnamefont {E.~L.}\ \bibnamefont
  {Hamilton}}, \bibinfo {author} {\bibfnamefont {C.~H.}\ \bibnamefont
  {Greene}},\ and\ \bibinfo {author} {\bibfnamefont {H.~R.}\ \bibnamefont
  {Sadeghpour}},\ }\href {https://doi.org/10.1088/0953-4075/35/10/102}
  {\bibfield  {journal} {\bibinfo  {journal} {J. Phys. B: At. Mol. Opt. Phys.}\
  }\textbf {\bibinfo {volume} {35}},\ \bibinfo {pages} {L199} (\bibinfo {year}
  {2002})}\BibitemShut {NoStop}%
\bibitem [{\citenamefont {Chibisov}\ \emph {et~al.}(2002)\citenamefont
  {Chibisov}, \citenamefont {Khuskivadze},\ and\ \citenamefont
  {Fabrikant}}]{Chibisov2002}%
  \BibitemOpen
  \bibfield  {author} {\bibinfo {author} {\bibfnamefont {M.~I.}\ \bibnamefont
  {Chibisov}}, \bibinfo {author} {\bibfnamefont {A.~A.}\ \bibnamefont
  {Khuskivadze}},\ and\ \bibinfo {author} {\bibfnamefont {I.~I.}\ \bibnamefont
  {Fabrikant}},\ }\href {https://doi.org/10.1088/0953-4075/35/10/101}
  {\bibfield  {journal} {\bibinfo  {journal} {J. Phys. B: At. Mol. Opt. Phys.}\
  }\textbf {\bibinfo {volume} {35}},\ \bibinfo {pages} {L193} (\bibinfo {year}
  {2002})}\BibitemShut {NoStop}%
\bibitem [{\citenamefont {Khuskivadze}\ \emph {et~al.}(2002)\citenamefont
  {Khuskivadze}, \citenamefont {Chibisov},\ and\ \citenamefont
  {Fabrikant}}]{Khuskivadze2002}%
  \BibitemOpen
  \bibfield  {author} {\bibinfo {author} {\bibfnamefont {A.~A.}\ \bibnamefont
  {Khuskivadze}}, \bibinfo {author} {\bibfnamefont {M.~I.}\ \bibnamefont
  {Chibisov}},\ and\ \bibinfo {author} {\bibfnamefont {I.~I.}\ \bibnamefont
  {Fabrikant}},\ }\href {https://doi.org/10.1103/PhysRevA.66.042709} {\bibfield
   {journal} {\bibinfo  {journal} {Phys. Rev. A}\ }\textbf {\bibinfo {volume}
  {66}},\ \bibinfo {pages} {042709} (\bibinfo {year} {2002})}\BibitemShut
  {NoStop}%
\bibitem [{\citenamefont {Greene}\ \emph {et~al.}(2006)\citenamefont {Greene},
  \citenamefont {Hamilton}, \citenamefont {Crowell}, \citenamefont {Vadla},\
  and\ \citenamefont {Niemax}}]{Greene2006}%
  \BibitemOpen
  \bibfield  {author} {\bibinfo {author} {\bibfnamefont {C.~H.}\ \bibnamefont
  {Greene}}, \bibinfo {author} {\bibfnamefont {E.~L.}\ \bibnamefont
  {Hamilton}}, \bibinfo {author} {\bibfnamefont {H.}~\bibnamefont {Crowell}},
  \bibinfo {author} {\bibfnamefont {C.}~\bibnamefont {Vadla}},\ and\ \bibinfo
  {author} {\bibfnamefont {K.}~\bibnamefont {Niemax}},\ }\href
  {https://doi.org/10.1103/PhysRevLett.97.233002} {\bibfield  {journal}
  {\bibinfo  {journal} {Phys. Rev. Lett.}\ }\textbf {\bibinfo {volume} {97}},\
  \bibinfo {pages} {233002} (\bibinfo {year} {2006})}\BibitemShut {NoStop}%
\bibitem [{\citenamefont {Giannakeas}\ \emph {et~al.}(2020)\citenamefont
  {Giannakeas}, \citenamefont {Eiles}, \citenamefont {Robicheaux},\ and\
  \citenamefont {Rost}}]{Giannakeas2020a}%
  \BibitemOpen
  \bibfield  {author} {\bibinfo {author} {\bibfnamefont {P.}~\bibnamefont
  {Giannakeas}}, \bibinfo {author} {\bibfnamefont {M.~T.}\ \bibnamefont
  {Eiles}}, \bibinfo {author} {\bibfnamefont {F.}~\bibnamefont {Robicheaux}},\
  and\ \bibinfo {author} {\bibfnamefont {J.~M.}\ \bibnamefont {Rost}},\ }\href
  {https://doi.org/10.1103/PhysRevA.102.033315} {\bibfield  {journal} {\bibinfo
   {journal} {Phys. Rev. A}\ }\textbf {\bibinfo {volume} {102}},\ \bibinfo
  {pages} {033315} (\bibinfo {year} {2020})}\BibitemShut {NoStop}%
\bibitem [{\citenamefont {Niederpr\"um}\ \emph {et~al.}(2015)\citenamefont
  {Niederpr\"um}, \citenamefont {Thomas}, \citenamefont {Manthey},
  \citenamefont {Weber},\ and\ \citenamefont {Ott}}]{Niederprum2015}%
  \BibitemOpen
  \bibfield  {author} {\bibinfo {author} {\bibfnamefont {T.}~\bibnamefont
  {Niederpr\"um}}, \bibinfo {author} {\bibfnamefont {O.}~\bibnamefont
  {Thomas}}, \bibinfo {author} {\bibfnamefont {T.}~\bibnamefont {Manthey}},
  \bibinfo {author} {\bibfnamefont {T.~M.}\ \bibnamefont {Weber}},\ and\
  \bibinfo {author} {\bibfnamefont {H.}~\bibnamefont {Ott}},\ }\href
  {https://doi.org/10.1103/PhysRevLett.115.013003} {\bibfield  {journal}
  {\bibinfo  {journal} {Phys. Rev. Lett.}\ }\textbf {\bibinfo {volume} {115}},\
  \bibinfo {pages} {013003} (\bibinfo {year} {2015})}\BibitemShut {NoStop}%
\bibitem [{\citenamefont {Gallagher}(1994)}]{Gallagher1994}%
  \BibitemOpen
  \bibfield  {author} {\bibinfo {author} {\bibfnamefont {T.~F.}\ \bibnamefont
  {Gallagher}},\ }\href {https://doi.org/10.1017/CBO9780511524530} {\emph
  {\bibinfo {title} {{Rydberg} Atoms}}},\ Cambridge Monographs on Atomic,
  Molecular and Chemical Physics\ (\bibinfo  {publisher} {Cambridge University
  Press},\ \bibinfo {year} {1994})\BibitemShut {NoStop}%
\bibitem [{\citenamefont {Clark}(1979)}]{Clark1979}%
  \BibitemOpen
  \bibfield  {author} {\bibinfo {author} {\bibfnamefont {C.~W.}\ \bibnamefont
  {Clark}},\ }\href {https://doi.org/10.1016/0375-9601(79)90127-0} {\bibfield
  {journal} {\bibinfo  {journal} {Phys. Lett. A}\ }\textbf {\bibinfo {volume}
  {70}},\ \bibinfo {pages} {295 } (\bibinfo {year} {1979})}\BibitemShut
  {NoStop}%
\bibitem [{\citenamefont {Fabrikant}(1986)}]{Fabrikant1986}%
  \BibitemOpen
  \bibfield  {author} {\bibinfo {author} {\bibfnamefont {I.~I.}\ \bibnamefont
  {Fabrikant}},\ }\href {https://doi.org/10.1088/0022-3700/19/10/021}
  {\bibfield  {journal} {\bibinfo  {journal} {J. Phys. B}\ }\textbf {\bibinfo
  {volume} {19}},\ \bibinfo {pages} {1527} (\bibinfo {year}
  {1986})}\BibitemShut {NoStop}%
\bibitem [{\citenamefont {Hölzl}(2019)}]{Holzl2019}%
  \BibitemOpen
  \bibfield  {author} {\bibinfo {author} {\bibfnamefont {C.}~\bibnamefont
  {Hölzl}},\ }\emph {\bibinfo {title} {Towards High $n$ Circular {Rydberg}
  States in Ultracold Atomic Gases}},\ \href@noop {} {Master's thesis},\
  \bibinfo  {school} {Universität Stuttgart} (\bibinfo {year}
  {2019})\BibitemShut {NoStop}%
\bibitem [{\citenamefont {Lorenzen}\ and\ \citenamefont
  {Niemax}(1983)}]{Lorenzen1983}%
  \BibitemOpen
  \bibfield  {author} {\bibinfo {author} {\bibfnamefont {C.-J.}\ \bibnamefont
  {Lorenzen}}\ and\ \bibinfo {author} {\bibfnamefont {K.}~\bibnamefont
  {Niemax}},\ }\href {https://doi.org/10.1088/0031-8949/27/4/012} {\bibfield
  {journal} {\bibinfo  {journal} {Physica Scripta}\ }\textbf {\bibinfo {volume}
  {27}},\ \bibinfo {pages} {300} (\bibinfo {year} {1983})}\BibitemShut
  {NoStop}%
\bibitem [{\citenamefont {Han}\ \emph {et~al.}(2006)\citenamefont {Han},
  \citenamefont {Jamil}, \citenamefont {Norum}, \citenamefont {Tanner},\ and\
  \citenamefont {Gallagher}}]{Han2006}%
  \BibitemOpen
  \bibfield  {author} {\bibinfo {author} {\bibfnamefont {J.}~\bibnamefont
  {Han}}, \bibinfo {author} {\bibfnamefont {Y.}~\bibnamefont {Jamil}}, \bibinfo
  {author} {\bibfnamefont {D.~V.~L.}\ \bibnamefont {Norum}}, \bibinfo {author}
  {\bibfnamefont {P.~J.}\ \bibnamefont {Tanner}},\ and\ \bibinfo {author}
  {\bibfnamefont {T.~F.}\ \bibnamefont {Gallagher}},\ }\href
  {https://doi.org/10.1103/PhysRevA.74.054502} {\bibfield  {journal} {\bibinfo
  {journal} {Phys. Rev. A}\ }\textbf {\bibinfo {volume} {74}},\ \bibinfo
  {pages} {054502} (\bibinfo {year} {2006})}\BibitemShut {NoStop}%
\bibitem [{\citenamefont {Goy}\ \emph {et~al.}(1982)\citenamefont {Goy},
  \citenamefont {Raimond}, \citenamefont {Vitrant},\ and\ \citenamefont
  {Haroche}}]{Goy1982}%
  \BibitemOpen
  \bibfield  {author} {\bibinfo {author} {\bibfnamefont {P.}~\bibnamefont
  {Goy}}, \bibinfo {author} {\bibfnamefont {J.~M.}\ \bibnamefont {Raimond}},
  \bibinfo {author} {\bibfnamefont {G.}~\bibnamefont {Vitrant}},\ and\ \bibinfo
  {author} {\bibfnamefont {S.}~\bibnamefont {Haroche}},\ }\href
  {https://doi.org/10.1103/PhysRevA.26.2733} {\bibfield  {journal} {\bibinfo
  {journal} {Phys. Rev. A}\ }\textbf {\bibinfo {volume} {26}},\ \bibinfo
  {pages} {2733} (\bibinfo {year} {1982})}\BibitemShut {NoStop}%
\bibitem [{\citenamefont {Weber}\ and\ \citenamefont
  {Sansonetti}(1987)}]{Weber1987}%
  \BibitemOpen
  \bibfield  {author} {\bibinfo {author} {\bibfnamefont {K.-H.}\ \bibnamefont
  {Weber}}\ and\ \bibinfo {author} {\bibfnamefont {C.~J.}\ \bibnamefont
  {Sansonetti}},\ }\href {https://doi.org/10.1103/PhysRevA.35.4650} {\bibfield
  {journal} {\bibinfo  {journal} {Phys. Rev. A}\ }\textbf {\bibinfo {volume}
  {35}},\ \bibinfo {pages} {4650} (\bibinfo {year} {1987})}\BibitemShut
  {NoStop}%
\bibitem [{\citenamefont {Eiles}(2018)}]{Eiles2018}%
  \BibitemOpen
  \bibfield  {author} {\bibinfo {author} {\bibfnamefont {M.~T.}\ \bibnamefont
  {Eiles}},\ }\href {https://doi.org/10.1103/PhysRevA.98.042706} {\bibfield
  {journal} {\bibinfo  {journal} {Phys. Rev. A}\ }\textbf {\bibinfo {volume}
  {98}},\ \bibinfo {pages} {042706} (\bibinfo {year} {2018})}\BibitemShut
  {NoStop}%
\bibitem [{\citenamefont {Karule}(1965)}]{Karule1965}%
  \BibitemOpen
  \bibfield  {author} {\bibinfo {author} {\bibfnamefont {E.}~\bibnamefont
  {Karule}},\ }\href
  {https://doi.org/https://doi.org/10.1016/0031-9163(65)91312-0} {\bibfield
  {journal} {\bibinfo  {journal} {Phys. Lett.}\ }\textbf {\bibinfo {volume}
  {15}},\ \bibinfo {pages} {137 } (\bibinfo {year} {1965})}\BibitemShut
  {NoStop}%
\bibitem [{\citenamefont {Engel}\ \emph {et~al.}(2019)\citenamefont {Engel},
  \citenamefont {Dieterle}, \citenamefont {Hummel}, \citenamefont {Fey},
  \citenamefont {Schmelcher}, \citenamefont {L\"ow}, \citenamefont {Pfau},\
  and\ \citenamefont {Meinert}}]{Engel2019}%
  \BibitemOpen
  \bibfield  {author} {\bibinfo {author} {\bibfnamefont {F.}~\bibnamefont
  {Engel}}, \bibinfo {author} {\bibfnamefont {T.}~\bibnamefont {Dieterle}},
  \bibinfo {author} {\bibfnamefont {F.}~\bibnamefont {Hummel}}, \bibinfo
  {author} {\bibfnamefont {C.}~\bibnamefont {Fey}}, \bibinfo {author}
  {\bibfnamefont {P.}~\bibnamefont {Schmelcher}}, \bibinfo {author}
  {\bibfnamefont {R.}~\bibnamefont {L\"ow}}, \bibinfo {author} {\bibfnamefont
  {T.}~\bibnamefont {Pfau}},\ and\ \bibinfo {author} {\bibfnamefont
  {F.}~\bibnamefont {Meinert}},\ }\href
  {https://doi.org/10.1103/PhysRevLett.123.073003} {\bibfield  {journal}
  {\bibinfo  {journal} {Phys. Rev. Lett.}\ }\textbf {\bibinfo {volume} {123}},\
  \bibinfo {pages} {073003} (\bibinfo {year} {2019})}\BibitemShut {NoStop}%
\bibitem [{\citenamefont {Sa\ss{}mannshausen}\ \emph
  {et~al.}(2015)\citenamefont {Sa\ss{}mannshausen}, \citenamefont {Merkt},\
  and\ \citenamefont {Deiglmayr}}]{Sassmannshausen2015}%
  \BibitemOpen
  \bibfield  {author} {\bibinfo {author} {\bibfnamefont {H.}~\bibnamefont
  {Sa\ss{}mannshausen}}, \bibinfo {author} {\bibfnamefont {F.}~\bibnamefont
  {Merkt}},\ and\ \bibinfo {author} {\bibfnamefont {J.}~\bibnamefont
  {Deiglmayr}},\ }\href {https://doi.org/10.1103/PhysRevLett.114.133201}
  {\bibfield  {journal} {\bibinfo  {journal} {Phys. Rev. Lett.}\ }\textbf
  {\bibinfo {volume} {114}},\ \bibinfo {pages} {133201} (\bibinfo {year}
  {2015})}\BibitemShut {NoStop}%
\bibitem [{\citenamefont {Miffre}\ \emph {et~al.}(2006)\citenamefont {Miffre},
  \citenamefont {Jacquey}, \citenamefont {Büchner}, \citenamefont {Tr\'enec},\
  and\ \citenamefont {Vigu\'e}}]{Miffre2006}%
  \BibitemOpen
  \bibfield  {author} {\bibinfo {author} {\bibfnamefont {A.}~\bibnamefont
  {Miffre}}, \bibinfo {author} {\bibfnamefont {M.}~\bibnamefont {Jacquey}},
  \bibinfo {author} {\bibfnamefont {M.}~\bibnamefont {Büchner}}, \bibinfo
  {author} {\bibfnamefont {G.}~\bibnamefont {Tr\'enec}},\ and\ \bibinfo
  {author} {\bibfnamefont {J.}~\bibnamefont {Vigu\'e}},\ }\href
  {https://doi.org/10.1140/epjd/e2006-00015-5} {\bibfield  {journal} {\bibinfo
  {journal} {Eur. Phys. J. D}\ }\textbf {\bibinfo {volume} {38}},\ \bibinfo
  {pages} {353–365} (\bibinfo {year} {2006})}\BibitemShut {NoStop}%
\bibitem [{\citenamefont {Holmgren}\ \emph {et~al.}(2010)\citenamefont
  {Holmgren}, \citenamefont {Revelle}, \citenamefont {Lonij},\ and\
  \citenamefont {Cronin}}]{Holmgren2010}%
  \BibitemOpen
  \bibfield  {author} {\bibinfo {author} {\bibfnamefont {W.~F.}\ \bibnamefont
  {Holmgren}}, \bibinfo {author} {\bibfnamefont {M.~C.}\ \bibnamefont
  {Revelle}}, \bibinfo {author} {\bibfnamefont {V.~P.~A.}\ \bibnamefont
  {Lonij}},\ and\ \bibinfo {author} {\bibfnamefont {A.~D.}\ \bibnamefont
  {Cronin}},\ }\href {https://doi.org/10.1103/PhysRevA.81.053607} {\bibfield
  {journal} {\bibinfo  {journal} {Phys. Rev. A}\ }\textbf {\bibinfo {volume}
  {81}},\ \bibinfo {pages} {053607} (\bibinfo {year} {2010})}\BibitemShut
  {NoStop}%
\bibitem [{\citenamefont {Amini}\ and\ \citenamefont
  {Gould}(2003)}]{Amini2003}%
  \BibitemOpen
  \bibfield  {author} {\bibinfo {author} {\bibfnamefont {J.~M.}\ \bibnamefont
  {Amini}}\ and\ \bibinfo {author} {\bibfnamefont {H.}~\bibnamefont {Gould}},\
  }\href {https://doi.org/10.1103/PhysRevLett.91.153001} {\bibfield  {journal}
  {\bibinfo  {journal} {Phys. Rev. Lett.}\ }\textbf {\bibinfo {volume} {91}},\
  \bibinfo {pages} {153001} (\bibinfo {year} {2003})}\BibitemShut {NoStop}%
\bibitem [{\citenamefont {Mitroy}\ \emph {et~al.}(2010)\citenamefont {Mitroy},
  \citenamefont {Safranova},\ and\ \citenamefont {Clark}}]{Mitroy2010}%
  \BibitemOpen
  \bibfield  {author} {\bibinfo {author} {\bibfnamefont {J.}~\bibnamefont
  {Mitroy}}, \bibinfo {author} {\bibfnamefont {M.~S.}\ \bibnamefont
  {Safranova}},\ and\ \bibinfo {author} {\bibfnamefont {C.~W.}\ \bibnamefont
  {Clark}},\ }\href {https://doi.org/10.1088/0953-4075/43/20/202001} {\bibfield
   {journal} {\bibinfo  {journal} {J. Phys. B: At. Mol. Opt. Phys.}\ }\textbf
  {\bibinfo {volume} {43}},\ \bibinfo {pages} {202001} (\bibinfo {year}
  {2010})}\BibitemShut {NoStop}%
\bibitem [{\citenamefont {Schlagm\"uller}\ \emph {et~al.}(2016)\citenamefont
  {Schlagm\"uller}, \citenamefont {Liebisch}, \citenamefont {Engel},
  \citenamefont {Kleinbach}, \citenamefont {B\"ottcher}, \citenamefont
  {Hermann}, \citenamefont {Westphal}, \citenamefont {Gaj}, \citenamefont
  {L\"ow}, \citenamefont {Hofferberth}, \citenamefont {Pfau}, \citenamefont
  {P\'erez-R\'{\i}os},\ and\ \citenamefont {Greene}}]{Schlagmuller2016x}%
  \BibitemOpen
  \bibfield  {author} {\bibinfo {author} {\bibfnamefont {M.}~\bibnamefont
  {Schlagm\"uller}}, \bibinfo {author} {\bibfnamefont {T.~C.}\ \bibnamefont
  {Liebisch}}, \bibinfo {author} {\bibfnamefont {F.}~\bibnamefont {Engel}},
  \bibinfo {author} {\bibfnamefont {K.~S.}\ \bibnamefont {Kleinbach}}, \bibinfo
  {author} {\bibfnamefont {F.}~\bibnamefont {B\"ottcher}}, \bibinfo {author}
  {\bibfnamefont {U.}~\bibnamefont {Hermann}}, \bibinfo {author} {\bibfnamefont
  {K.~M.}\ \bibnamefont {Westphal}}, \bibinfo {author} {\bibfnamefont
  {A.}~\bibnamefont {Gaj}}, \bibinfo {author} {\bibfnamefont {R.}~\bibnamefont
  {L\"ow}}, \bibinfo {author} {\bibfnamefont {S.}~\bibnamefont {Hofferberth}},
  \bibinfo {author} {\bibfnamefont {T.}~\bibnamefont {Pfau}}, \bibinfo {author}
  {\bibfnamefont {J.}~\bibnamefont {P\'erez-R\'{\i}os}},\ and\ \bibinfo
  {author} {\bibfnamefont {C.~H.}\ \bibnamefont {Greene}},\ }\href
  {https://doi.org/10.1103/PhysRevX.6.031020} {\bibfield  {journal} {\bibinfo
  {journal} {Phys. Rev. X}\ }\textbf {\bibinfo {volume} {6}},\ \bibinfo {pages}
  {031020} (\bibinfo {year} {2016})}\BibitemShut {NoStop}%
\bibitem [{\citenamefont {Kanungo}\ \emph {et~al.}(2020)\citenamefont
  {Kanungo}, \citenamefont {Whalen}, \citenamefont {Lu}, \citenamefont
  {Killian}, \citenamefont {Dunning}, \citenamefont {Yoshida},\ and\
  \citenamefont {Burgd\"orfer}}]{Kanungo2020}%
  \BibitemOpen
  \bibfield  {author} {\bibinfo {author} {\bibfnamefont {S.~K.}\ \bibnamefont
  {Kanungo}}, \bibinfo {author} {\bibfnamefont {J.~D.}\ \bibnamefont {Whalen}},
  \bibinfo {author} {\bibfnamefont {Y.}~\bibnamefont {Lu}}, \bibinfo {author}
  {\bibfnamefont {T.~C.}\ \bibnamefont {Killian}}, \bibinfo {author}
  {\bibfnamefont {F.~B.}\ \bibnamefont {Dunning}}, \bibinfo {author}
  {\bibfnamefont {S.}~\bibnamefont {Yoshida}},\ and\ \bibinfo {author}
  {\bibfnamefont {J.}~\bibnamefont {Burgd\"orfer}},\ }\href
  {https://doi.org/10.1103/PhysRevA.102.063317} {\bibfield  {journal} {\bibinfo
   {journal} {Phys. Rev. A}\ }\textbf {\bibinfo {volume} {102}},\ \bibinfo
  {pages} {063317} (\bibinfo {year} {2020})}\BibitemShut {NoStop}%
\bibitem [{\citenamefont {Geppert}\ \emph {et~al.}(2020)\citenamefont
  {Geppert}, \citenamefont {Althön}, \citenamefont {Fichtner},\ and\
  \citenamefont {Ott}}]{Geppert2020}%
  \BibitemOpen
  \bibfield  {author} {\bibinfo {author} {\bibfnamefont {P.}~\bibnamefont
  {Geppert}}, \bibinfo {author} {\bibfnamefont {M.}~\bibnamefont {Althön}},
  \bibinfo {author} {\bibfnamefont {D.}~\bibnamefont {Fichtner}},\ and\
  \bibinfo {author} {\bibfnamefont {H.}~\bibnamefont {Ott}},\ }\href@noop {}
  {\bibinfo {title} {Diffusive-like redistribution in state-changing collisions
  between {Rydberg} atoms and ground state atoms}} (\bibinfo {year} {2020}),\
  \Eprint {https://arxiv.org/abs/2012.11485} {arXiv:2012.11485
  [physics.atom-ph]} \BibitemShut {NoStop}%
\end{thebibliography}
